\documentstyle{aca}

\title{Extinction in the Galaxy from surface brightnesses of ESO-LV galaxies:
 testing 'standard' extinction maps}

\author[J. Cho{\l}oniewski, E. A. Valentijn]
       {Jacek Cho{\l}oniewski,$^1$ Edwin A. Valentijn$^2$\\
       $^1$Astronomical Observatory of Warsaw University,
       Aleje Ujazdowskie 4, PL-00478 Warsaw, Poland,
       e-mail: astro@estymator.com.pl\\
       $^2$Kapteyn Institute, P. O. Box 800, NL-9700 AV Groningen, The Netherlands,
       e-mail: valentyn@astro.rug.nl}

\begin{document}

\maketitle

\begin{abstract}
The relative extinction in the Galaxy computed 
with our new method 
(Cho{\l}oniewski \& Valentijn 1999, CV)
is compared with three patterns:
Schlegel, Finkbeiner \& Davis (1998, SFD),
Burstein \& Heiles (1978, BH) and the cosecans law.
It is shown that extinction of SFD is more reliable
then that of BH since it stronger correlates with our new
extinction.
The smallest correlation coeffcient have been obtained
for the cosecans law.
Linear regression analysis show that SFD overestimate the extinction
by a factor of 1.4.

Our results clearly indicate that there is non-zero extinction
at the Galactic South pole and that the extinction near the Galactic
equator ($|b|<40^o$) is significantly larger in the Southern hemisphere
than in the Northern.

\end{abstract}


\section{Introduction}

We have introduced in our previous paper 
(Cho{\l}oniewski \& Valentijn 1999, hereafter CV) a
new method for the determination of the extinction in our Galaxy.
The method uses surface brightnesses of external galaxies in the B and
R bands
as listed in {\it The Surface Photometry Catalogue of the ESO-Uppsala
Galaxies} (Lauberts \& Valentijn, 1989, hereafter ESO-LV).
The first draft of this method has been 
published in our earlier paper (Cho{\l}oniewski \& Valentijn 1991).

The main purpose of the present paper is to compare our derived extinction
values with recently published maps of extinction by
Schlegel, Finkbeiner \& Davis (1998), hereafter SFD, and with the
frequently used map of  Burstein \& Heiles (1978), hereafter BH
(see also Burstein \& Heiles 1982).

\section{The method}

Our extinction determination (fully described in CV) employs
 the surface brightnesses of
external galaxies in the B and R bands: $\mu_B$, $\mu_R$.
Basically, our method produces the relative extinction compared to an
overall mean - extinction with an {\it unknown zero-point}.
% ** I dropped next sentence
%Such relative extinction can be positive and negative as well.
The formula for the relative extinction (in B band) is simply a linear
combination of $\mu_B$ and $\mu_R$:

\begin{equation}
A_B = \frac{\mu_B - s \,\, \mu_R}{1 - r \,\, s} \, - \, c .
\end{equation}
In order to use equation (1) one
has to know three parameters: $r$, $s$ and $c$.

The parameter $r$ describes the ratio of extinction in the R and the B band
($r=A_R/A_B$) and is assumed to be constant.
Its recent literature value is 0.61 (see CV for
references) while we derived in CV two new
estimates: 0.62 and 0.64.
As a reasonable compromise we adopt throughout this
paper $r=0.62$. 

The inverse of the parameter $s$ describes the slope of the
linear relation between surface brightnesses $\mu_B$
and $\mu_R$.
The $c$ parameter is introduced in order to maintain
zero-point issues.
Both parameters $s$ and $c$ depend on morphological type $T$,
so they have been computed (using equations 7 and 10 in CV)
separately for every morphological type.

In this paper we will consider several different subsamples.
Extinction within every such subsample has been computed using 
the set of values $s(T)$ and $c(T)$ obtained from the same
set of data.
The $s(T)$ and $c(T)$ coefficients for the most important two subsamples
used in ths paper are in Table 1.

\begin{table}
 \caption{Coefficients $s(T)$ and $c(T)$ used for computing relative
extinction in this paper according to equation 1}
\begin{tabular}{@{}rrrrr@{}}
    &\multicolumn{2}{c}{sample "A"} &\multicolumn{2}{c}{sample "B"} \\
    &        &        &        &        \\
  T &   s(T) &   c(T) &   s(T) &   c(T) \\
    &        &        &        &        \\
 -5 &  0.826 & 10.706 &  0.842 & 10.233 \\
 -4 &  0.955 &  6.263 &  0.997 &  4.199 \\
 -3 &  0.915 &  7.619 &  0.932 &  6.927 \\
 -2 &  0.893 &  8.295 &  0.894 &  8.306 \\
 -1 &  0.785 & 11.597 &  0.772 & 12.060 \\ 
  0 &  0.768 & 12.060 &  0.840 & 10.129 \\
  1 &  0.728 & 12.972 &  0.737 & 12.885 \\
  2 &  0.691 & 13.903 &  0.681 & 14.166 \\
  3 &  0.670 & 14.223 &  0.668 & 14.321 \\
  4 &  0.652 & 14.589 &  0.650 & 14.663 \\
  5 &  0.635 & 14.979 &  0.637 & 14.943 \\
  6 &  0.641 & 14.943 &  0.697 & 13.627 \\
  7 &  0.717 & 13.163 &  0.714 & 13.251 \\
  8 &  0.731 & 12.844 &  0.721 & 13.181 \\
  9 &  0.744 & 12.656 &  0.784 & 11.446 \\
 10 &  0.795 & 11.054&  0.806 & 10.675
\end{tabular}
\end{table}

We use in this paper the extinction
in B band as described in equation (1) and denote it
as $A_B(CV)$.

\section{The samples}
For our analysis we use {\it surface brightnesses at the radius of
 half total B light}
in B and R bands from ESO-LV.
We exclude from the sample those galaxies which have morphological 
classifications suspected to be extinction dependent (marked in ESO-LV
with $T_{flag}$ equal to 4). We also reject galaxies which have
excessive (probably wrong) colours: $\mu_B - \mu_R$ smaller than zero
or greater than 2.1. 

Since we focus in this paper on the comparison of our extinction
with SFD and BH data, we reject additionally those galaxies for
which BH or SFD extinction values are not available.

The largest complete subsample of galaxies of ESO-LV 
can be made by selecting galaxies
which have
{\it visual apparent diameter} $D_{org}$
greater or equal to 60 arcsec. 
It contains, after applying the rejections described above, 7974 galaxies
(sample "A").

As mentioned in CV, the completeness limit
of  ESO-LV galaxies is  morphological type dependent and
a universal criterion, valid for every morphological type,
is selecting galaxies with a visual diameter limit larger than 100 arcsec. 
We have used such more restricted sample for all the 
computations presented in CV. We use it also here as sample "B" (after
rejections described above).

Results described in Sections 4 and 5 show that more rigorousely
defined sample "B"
produces slightly more accurate extinction {\it per galaxy}
than the sample "A". 
However the sample "A" is three times larger than the sample "B"
what, at least in part, compensate its slightly larger dispersion.

Table 2 contains a summary of the  definitions of sample "A"
and sample "B".

Since the ESO-LV galaxy catalogue covers the Southern sky 
($\delta < -17.5^o$) our analysis refers to
this part of the hemisphere.

\begin{table}
 \caption{Definitions of the galaxy samples.}
 \begin{tabular}{@{}ll@{}}
 \multicolumn{2}{l}{General conditions:}\\ \\
 \multicolumn{2}{l}{$A_B(BH)$ - present}\\
 \multicolumn{2}{l}{$A_B(SFD)$ - present}\\
 \multicolumn{2}{l}{$T_{flag} \neq 4$}\\
 \multicolumn{2}{l}{$0 \leq \mu_B - \mu_R \leq 2.1$}\\ \\
  sample A & sample B \\ \\
  $D_{org} \ge 60 \, arcsec$ & $D_{org} \ge 100 \, arcsec$ \\ \\
  N=7974 & N=2450
 \end{tabular}
\end{table}

\section{The accuracy}

Equation (1) represents an estimator of the foreground relative 
extinction for an individual external
galaxy. It is important to know its uncertainty (standard error).
In order to obtain this, we have divided the whole sky
into squares with size $\Delta$ degrees and computed the average
variance of the extinction inside these squares using the formula given by 
the so called one-way analysis of the variance theory (see e. g. Fish 1962).
This average variance reflects the variations of the true extinction inside
the squares with size $\Delta$ and the standard error of the extinction
estimator expressed in equation (1).
So, when the size of the squares $\Delta$ tends to zero 
the average variance of the extinction should tend to
the standard error of the extinction produced by our method.

Figure 1 shows the average variance of extinction as a function
of $\Delta$. The minimum value for $\Delta$ which we applied was
one degree. For smaller values of $\Delta$ a too large fraction of studied
squares contain only one galaxy (such squares can not
be taken into account in the one-way analysis of variance).

As we expect, the average variance has minimum for the
smallest applied value of $\Delta$ (one degree). 
For sample "A" the standard error
of our extinction in $A_B$ is 0.43 magnitude, while for 
sample "B" this is 0.40 magnitude,
which corresponds to a standard error in $E(B-V)$ 
of approximately 0.10 magnitude (for
$A_B / E(B-V) = 4.3$).
Sample "B" produces relative extinction with slightly higher
accuracy than sample "A".

BH used for the calibration of their extinction map B-V colours
of 131 globular clusters and RR Lyrae stars. 
SFD used for the calibration of their extinction map B-R colours
of 106 brightest cluster ellipticals and B-V colours of
389 elliptical galaxies with measured $Mg_2$ index (505 objects
in total).

Both BH and SFD report that their calibrators show a
residual scatter in B-V, with respect to calibration regression line,
of approximately 0.03 magnitude.

Our extinction estimator, when applied to photographically
measured surface brightnesses of galaxies in two bands as listed
in the  ESO-LV catalogue, has three times larger standard
error than the calibrators used by BH and SFD. 
This is an important disadvantage (at least as long as we apply it
to ESO-LV photometrical data).
But there is one important advantage of our extinction estimator -
it can be applied to many more objects since
we can apply the method, at present, to 7974 galaxies (sample "A") 
from the ESO-LV
catalogue.

\section{The correlation}

We have computed the Pearson, Spearman and Kendall
correlation coefficients (see Press, Teukolsky, Vetterling \& 
Flannery 1992 for definitions and software) between our extinction 
and extinction given by SFD and BH and
for the cosecans law: 

\begin{equation}
A_B = A_0 \, \csc|b| ,
\end{equation}
where $b$ denotes Galactic latitude.
The computation have been performed for
sample "A" and sample "B" (see Table 3). 
The parameters $s(T)$ and $c(T)$ have been computed separately for every
sample (see Table 1).

\begin{table}
 \caption{Pearson, Spearman and Kendall correlation coefficients 
between $A_B(CV)$ extinction estimate and three other extinction estimates: SFD,
BH and csc(b).}
\begin{tabular}{@{}llll@{}}
       & Pearson & Spearman & Kendall \\ \\
\multicolumn{4}{l}{Sample "A" }\\ \\
SFD    &  0.276 & 0.234 & 0.158 \\
BH     &  0.247 & 0.219 & 0.148 \\
csc(b) &  0.202 & 0.188 & 0.127 \\ \\ 
\multicolumn{4}{l}{Sample "B" }\\ \\
SFD    & 0.327 & 0.304 & 0.208 \\
BH     & 0.293 & 0.282 & 0.192 \\
csc(b) & 0.238 & 0.252 & 0.170 
\end{tabular}
\end{table}

All three correlation coefficients for both samples are the largest for 
SFD extinction and the smallest for the cosecans law. The extinction of BH is
always between these two extreme results. 

The coefficients are generally higher for sample "B" than for sample "A"
what suggests that sample "B" produces more accurate extinction than
sample "A".

Since the sample "B" is a subsample of the sample "A" the
correlation coefficients for "A" and "B" are not statistically independent.
In order to produce a set of statistically independent correlations
we divide the sample "A" into six subsamples:

\begin{enumerate}
\item $60 \, arcsec \leq D_{org} < 80 \, arcsec$ (N=3876)
\item $80 \, arcsec \leq D_{org} < 100 \, arcsec$ (N=1648)
\item $100 \, arcsec \leq D_{org} < 120 \, arcsec$ (N=718)
\item $120 \, arcsec \leq D_{org} < 140 \, arcsec$ (N=647)
\item $140 \, arcsec \leq D_{org} < 160 \, arcsec$ (N=298)
\item $160 \, arcsec \leq D_{org}$ (N=787) ,
\end{enumerate}
and compute the parameters $s(T)$ and $c(T)$ separately for every 
subsample. As a result of this procedure we have six statistically
independent correlation coefficients - see Figs 2, 3 and 4 for
results. As before, the correlations are the largest for SFD, smaller
for BH and the smallest for cosecans law.

\section{Correction of the BH's extinction zero point}

The formulae of BH produces for some regions of the sky
extinction less than zero (for the samples analyzed in this
paper the minimum value of $A_B(BH)$ extinction is -0.12 magnitude). 
In spite of BH's instructions
to set these values to zero we have actually used these negative values and
found that $A_B(CV)$ is for them significantly smaller
than for $A_B(BH) \approx 0$ (see upper panel of Figs 7 and 8).
This means that the BH extinction (in the  B band) 
was underestimated by 0.12 magnitude - just the absolute
value of the minimum value of $A_B(BH)$ extinction. This is in 
approximate agreement
with SFD who discovered  a similar offset of 0.09 magnitude.
We use in the regression analysis presented in the next
Section the corrected BH's extinction:
$A_B(BH)_C \, = \, A_B(BH) + 0.12$
instead of $A_B(BH)$ itself.

\section{The linear regression}

In the ideal case there would be linear dependence with slope equal
to one between our relative extinction 
estimate (CV) and the extinction of SFD and BH (corrected).

Since our extinction values are relative, with an arbitrary
zero-point, the constant term of this linear dependence should not be
equal to zero. The constant, multiplied by $-1$, should be added to
our {\it relative} extinction to transform it to the {\it absolute} extinction.

We have fitted the straight lines (using the least squares method) taking as
independent variables extinction of SFD and BH and as dependent variable
our estimate of relative extinction (CV). 
We have found the following regression coefficients for sample "A":

\begin{equation}
A_B(CV) = 0.662\, (\pm 0.028) A_B(SFD) - 0.189\, (\pm 0.010)
\end{equation}
\begin{equation}
A_B(CV) = 0.555\, (\pm 0.026) A_B(BH)_C - 0.176\, (\pm 0.010)
\end{equation}
and for sample "B":
\begin{equation}
A_B(CV) = 0.741\, (\pm 0.043) A_B(SFD) - 0.225\, (\pm 0.016)
\end{equation}
\begin{equation}
A_B(CV) = 0.608\, (\pm 0.040) A_B(BH)_C - 0.206\, (\pm 0.017)
\end{equation}
where in brackets $1\sigma$ errors are given. 
The differences between slopes and constant terms computed for sample 
"A" and "B" are only marginally larger than the combined
errors. 
In the forthcoming we use their averages.

A graphical
presentations of the regression lines are shown in Figs 5-8.
In order to investigate whether the postulate about linear
dependence is valid, the regression lines are shown together with
row data (lower panels) and with averages of $A_B(CV)$ computed
for 0.05 magnitude bins of $A_B(SFD)$ and $A_B(BH)$ (upper panels).

The slopes of the regression lines are definitely less than
one: for SFD the slope is circa 0.7 while for BH it is circa 0.6. 
The constant terms for both SFD and BH  are approximately the 
same: -0.20 magnitude. This
defines the zero point of our results: adding 0.20 magnitude to
our {\it relative} extinction transforms it to the {\it absolute} extinction.

When we introduce our absolute extinction into equations 3 - 6, 
the constant terms in all 4 equations gets close to zero,
now setting the relation between our (absolute)
extinction and SFD's and BH's extinction. Thus we conclude:
 SFD overestimate extinction by a
factor of $0.7^{-1} \approx 1.4$ while BH by a factor of 
$0.6^{-1} \approx 1.7$.

Up to now five papers reported similar results, namely: that SFD
extinction is about 1.4 times too large.
Stanek (1998b) obtained, using colours of low galactic latitude
globular clusters, that the overestimation factor is 1.35 (see
also Stanek 1998a).
Arce \& Goodman (1999a, 1999b) analyzed extinction in the Taurus
region using four independent methods and
found that the factor is 1.3 - 1.5.
Gonzalez, Fruchter \& Dirsch (1999) obtain extinction in the
small field around GRB 970228. They found using two methods
$A_V = 0.55$ while $A_V(SFD) = 0.78$. The ratio of these two
numbers give again a factor 1.4. Our finding that SFD overestimate 
extinction is also supported by Ivans et al. (1999) and von Braun \& Mateo
(2001).

Some kind of doubt about the correctness of the calibration is also in original SFD paper
where we can find as the comment to their Fig. 6 the statement that "A slight trend in 
the residuals is evident for both BH and DIRBE/IRAS corrections, in the sense that the
highest reddening values appear to be overestimated". 

Inspecting Figs 5-8 demonstrate a general linear
dependence between our relative extinction estimate and the
extinction of SFD and BH. 
However, some minor, but statistically significant, deviations
are clearly present, especially for $A_B(SFD)$ and  $A_B(BH)$
less than 0.5 magnitude.

\section{Galactic latitude dependence}

Unfortunately, we can not use our method of extinction determination,
as applied to the ESO-LV galaxy catalogue, to produce a new 
high resolution extinction
map and to compare it with SFD and BH maps. 
This is because our data are too sparse (less than one galaxy per square
degree) and their accuracy is too low 
($\sigma (E(B-V)) \approx 0.10$ per galaxy)
mainly because of instrinsic scatter of the surface brightness of galaxies.

But the quality of our data is good enough to evaluate their galactic
latitude dependence.
We show this dependence for our absolute extinction 
(equal to the relative extinction, as defined in equation 1, 
plus 0.20 magnitude) together with
SFD, corrected BH and the cosecans low - see Fig. 9.
All four solutions have been
averaged inside 5 degrees galactic latitude bins in the
points where ESO-LV galaxies are (sample "A").
We use $A_o=0.10$ for the cosecans law (equation 3) to match
as close as possible to our solution.

Fig. 9 shows that our absolute extinction is always greater than zero 
confirming the value of the zero-point computed in Section 7.

As one can see in Fig. 9 our solution as well as SFD and BH 
mimic quite well the classic cosecans law and indicate
considerable extinction near the galactic pole.
The differences between our results and SFD and BH are the largest
near the galactic equator  and 
at the northern hemisphere.

Since we find that both SFD and the (corrected) BH extinction standards
overestimate Galactic extinction by a factor 1.4 and 1.7 respectively,
we show the same data as in Fig.9 but with SFD and corrected BH
extinction divided by these factors 
(or, equivalently, multiplied by 0.7 and 0.6 respectively) -
see Fig. 10. The better agreement of our
solution with the rescaled SFD and BH data is evident, especially
near the Galactic equator ($b \approx 0$), confirming
the need for rescaling SFD and BH extinction.
But even for the rescaled data statistically significant
differences between our results and SFD and BH, although
smaller, still exist. Especially our data exhibit a South - North
asymmetry near the galactic equator ($|b| < 40^o$) with more
extinction in the Southern Galactic hemisphere which is
not visible in SFD and BH data.

\section{Discussion}

Our extinction estimator (introduced in CV) definitely
indicates that the new extinction map of SFD is more
realiable than the old one of BH. The historical
cosecans law is in this competition on the last place.
This result has been obtained using six separate, statistically
independent, sets of data.

The superiority of the SFD extinction map over the BH map have been 
demonstrated by
computing correlation coefficients with our extinction
results. 
However, the correlation coefficients are not sensitive to the
amplitude of variation of the input data nor its zero-point.
In order allow for further, more specific, comparisons we
have performed a linear regression analysis between
our relative extinction and SFD's and BH's extinction.
This analysis provides us the zero-point of
our extinction (equal to 0.20 magnitude) which allows us to
transform our relative extinction into absolute extinction.
We have showed that, in comparison to our absolute extinction,
SFD overestimate extinction by a factor of 1.4. 
This is in agreement with five other
authors and considerably changes our view on the amplitude of the 
Galactic extinction.

Superiority of SFD over BH reddening map, as we report
in this paper, does not mean that the first is an ideal result -
we have discovered some significant differences between our
extinction and SFD when analyzing their Galactic
latitude dependence even after correcting SFD extinction by
dividing it by 1.4 factor. 
One possible source of these differences
is that SFD assume that the extinction is strictly
proportional to the dust column density what need not be
true - the extinction to dust ratio can vary across the sky and
may also depend on the wavelength.

Our results are important for creating reliable extinction
standards and demonstrate the correctness and usefulness of our 
extinction estimator.
They can be also regarded as a stimulus for applying our extinction
estimator to other, larger and more accurate
galaxy catalogues, particularly thoose expected from  
new wide field imaging surveys,

Future applications of our method to other databases may produce maps
with a courser resolution.  At present the method applied to ESO-LV data
is unable to generate the extinction map with both sufficient resolution
and accuracy.  Our present method  should be rather taken as new extinction
calibrator. But using the method to
larger and more precise galaxy catalogues may result into stand-alone
high resolution extinction maps.

\section*{Acknowledgments}

The part of this work was done during the stay of the authors
at European Southern Observatory (Garching, Germany).
The authors thanks Andris Lauberts for
collaboration about the project.

\clearpage
\begin{figure*}
\vspace{220mm}
 \includegraphics{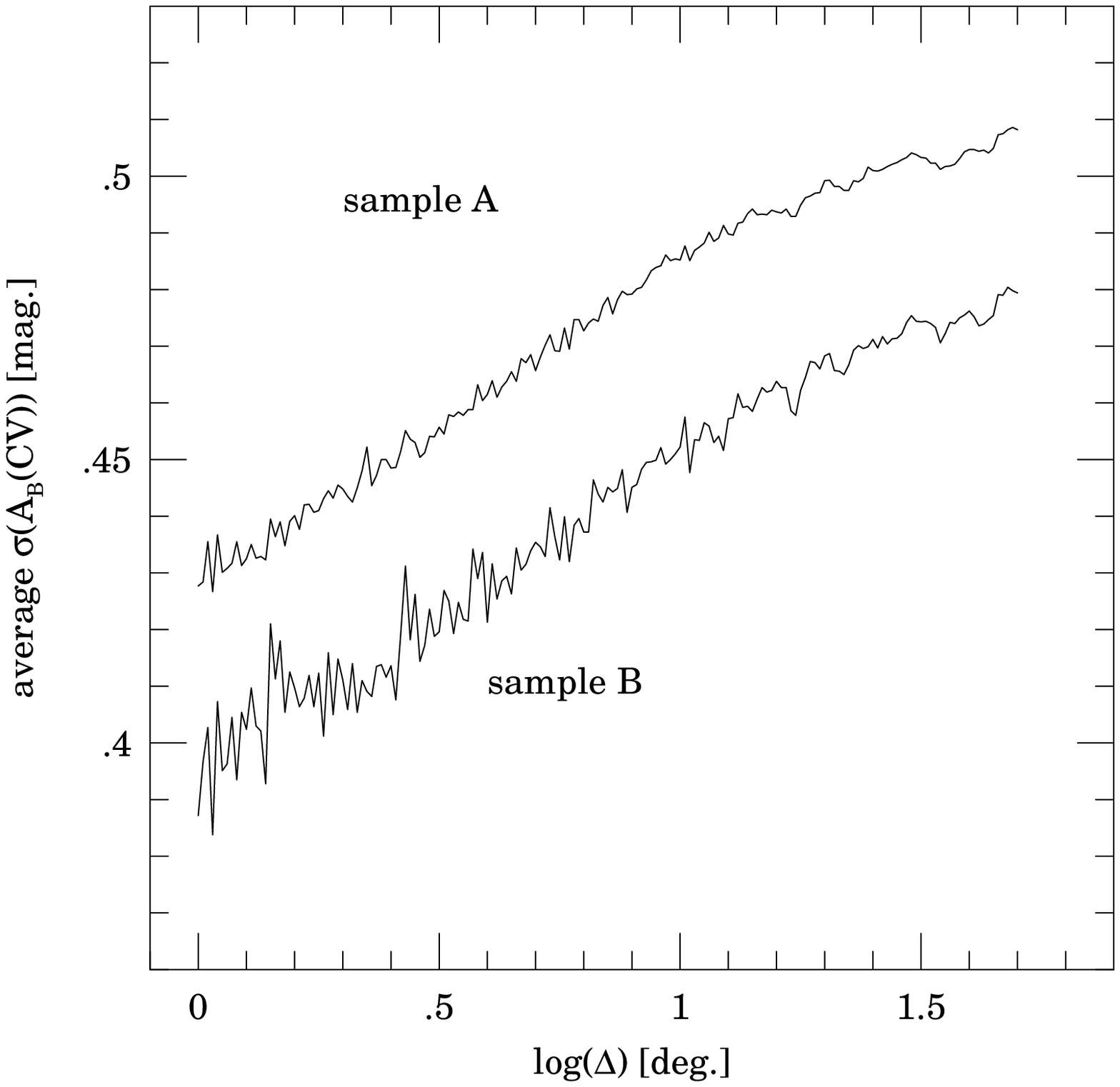}
 \caption{
The average variance of extinction $A_B(CV)$ inside $\Delta$
degrees squares on the sky as a function of the size ($\Delta$) of these
squares. The minimum variance (at one degree: $log(\Delta)=0$)
represent the standard deviation of the estimator of relative
extinction used in this paper.
}
\end{figure*}

\clearpage
\begin{figure*}
\vspace{220mm}
 \includegraphics{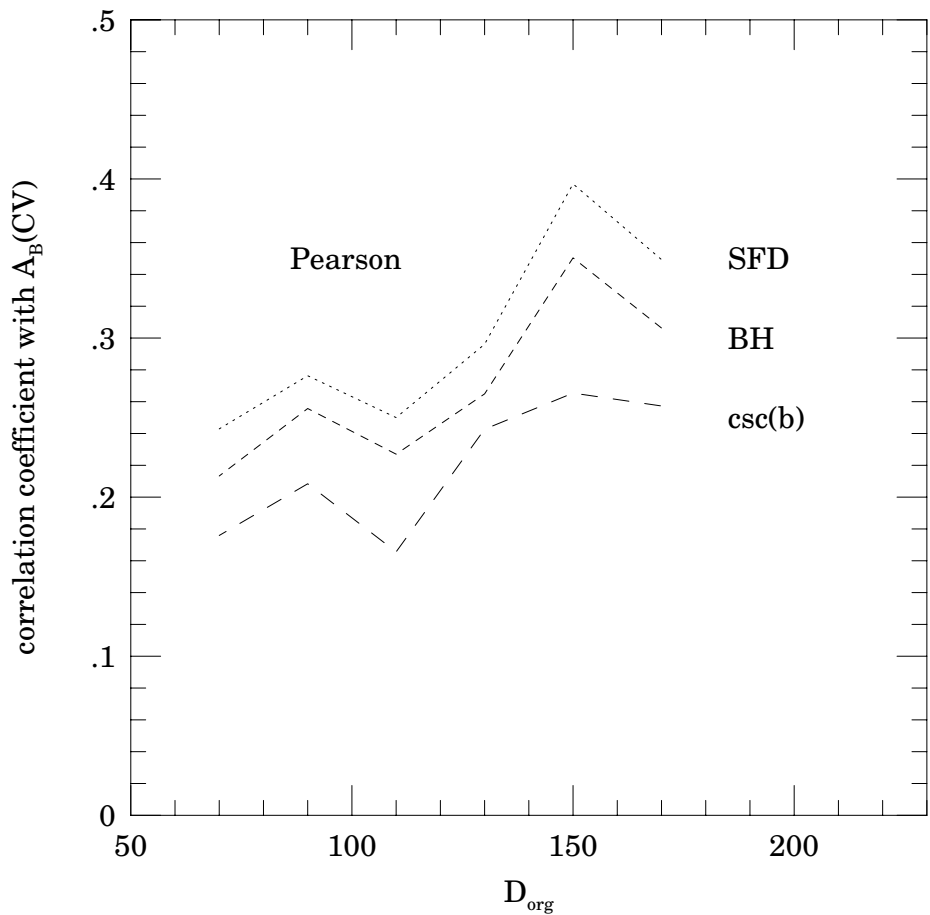}
 \caption{
Pearson correlation coefficient between our $A_B(CV)$ extinction and
the extinction of SFD and BH and the cosecans law computed for six
subsamples defined using visual diameter $D_{org}$ (see Section 5 for
details)
}
\end{figure*}

\clearpage
\begin{figure*}
\vspace{220mm}
 \includegraphics{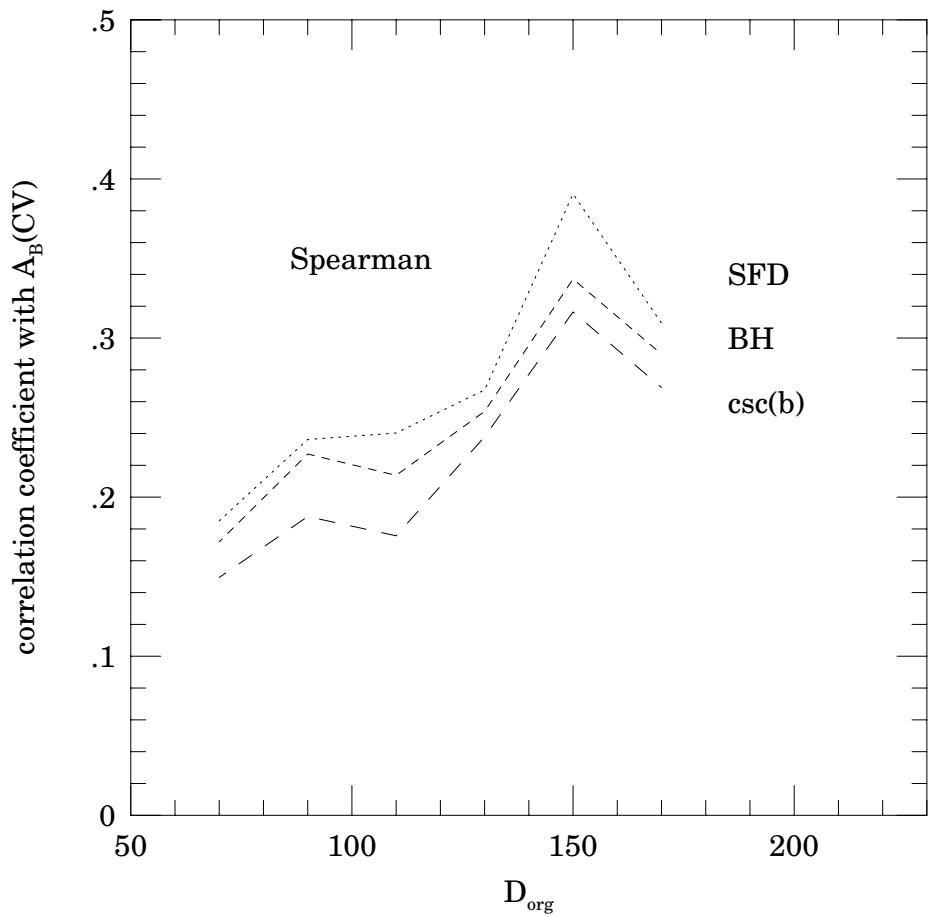}
 \caption{
The same as Fig. 2 but for Spearman correlation coefficient.
}
\end{figure*}

\clearpage
\begin{figure*}
\vspace{220mm}
 \includegraphics{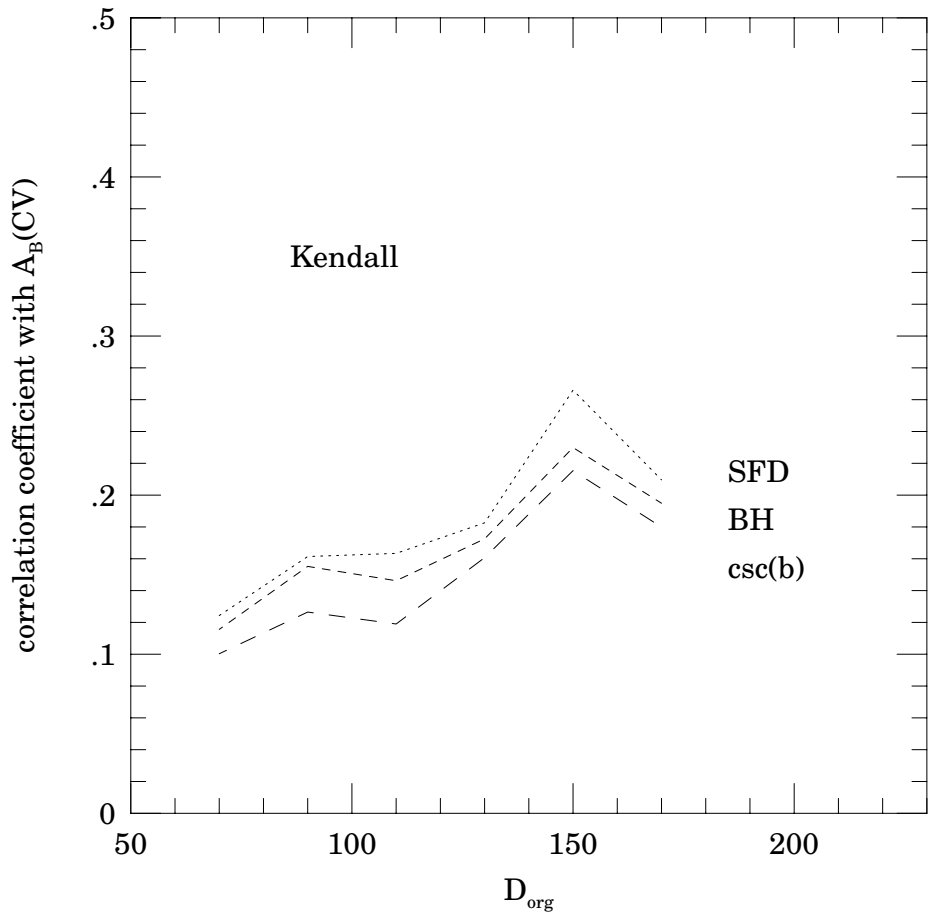}
 \caption{
The same as Fig. 3 but for Kendall correlation coefficient.
}
\end{figure*}

\clearpage
\begin{figure*}
\vspace{220mm}
 \includegraphics{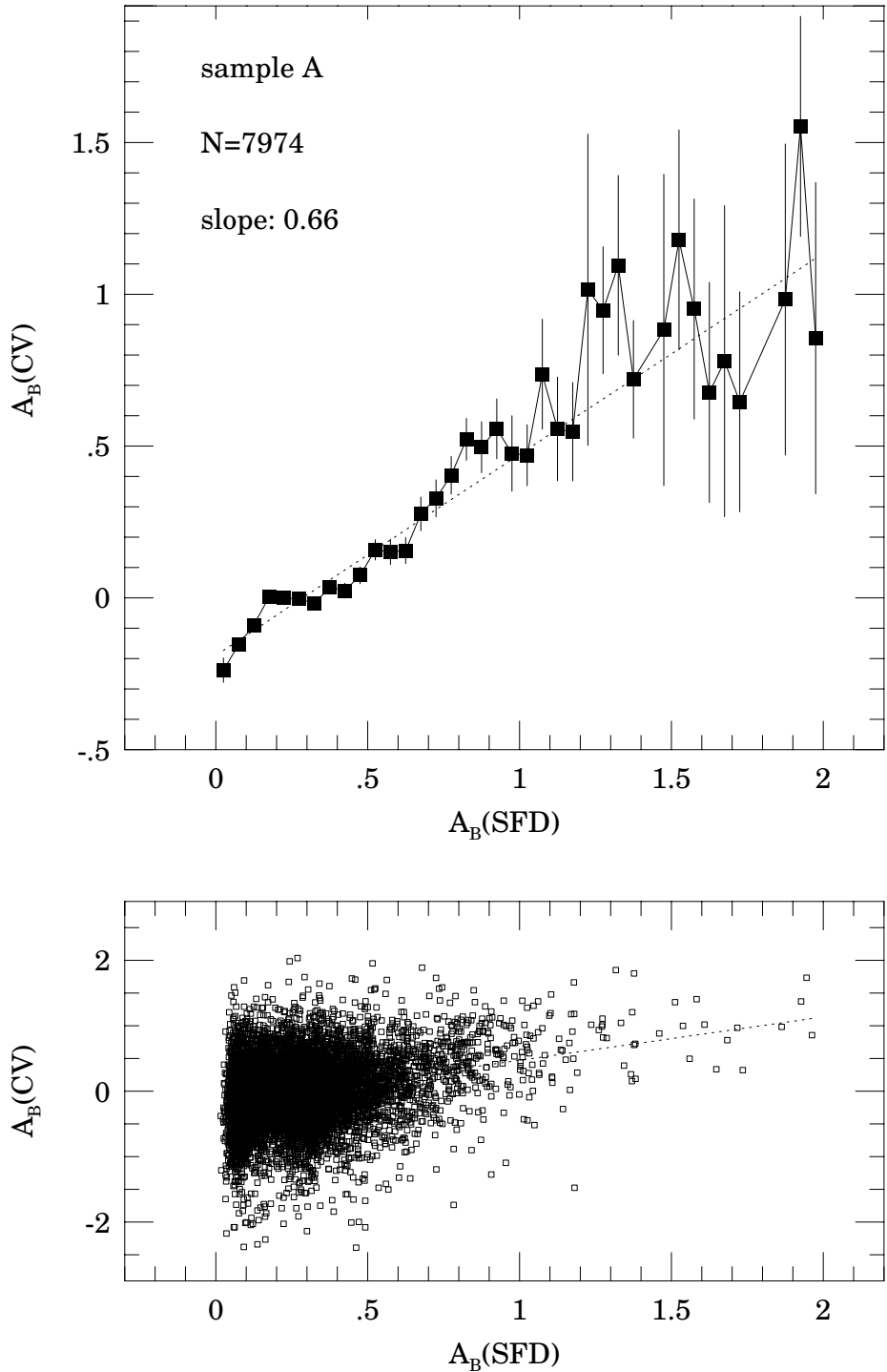}
 \caption{
The dependence between the extinction of SFD $A_B(SFD)$ and our
relative extinction $A_B(CV)$.
Upper panel shows the averaged data inside 0.05 magnitude bins.
Error bars represent standard deviation ($1\sigma$).
Lower panel shows raw data.
The data are taken from sample "A" (see text).
Dotted straight lines visible on both panels represent the least squares fit.
}
\end{figure*}

\clearpage
\begin{figure*}
\vspace{220mm}
 \includegraphics{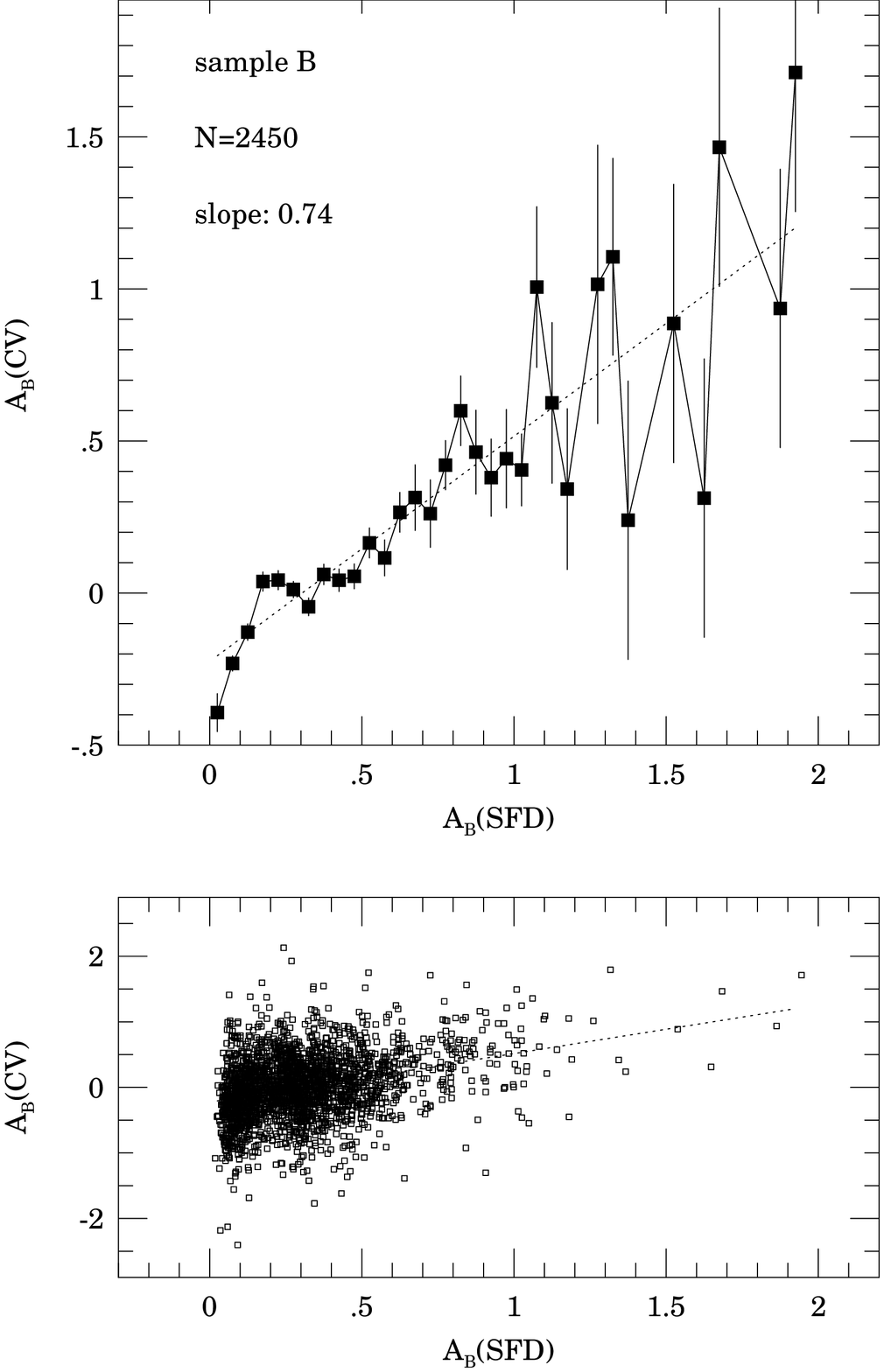}
 \caption{
The same as Fig. 5 but for sample "B".
}
\end{figure*}

\clearpage
\begin{figure*}
\vspace{220mm}
 \includegraphics{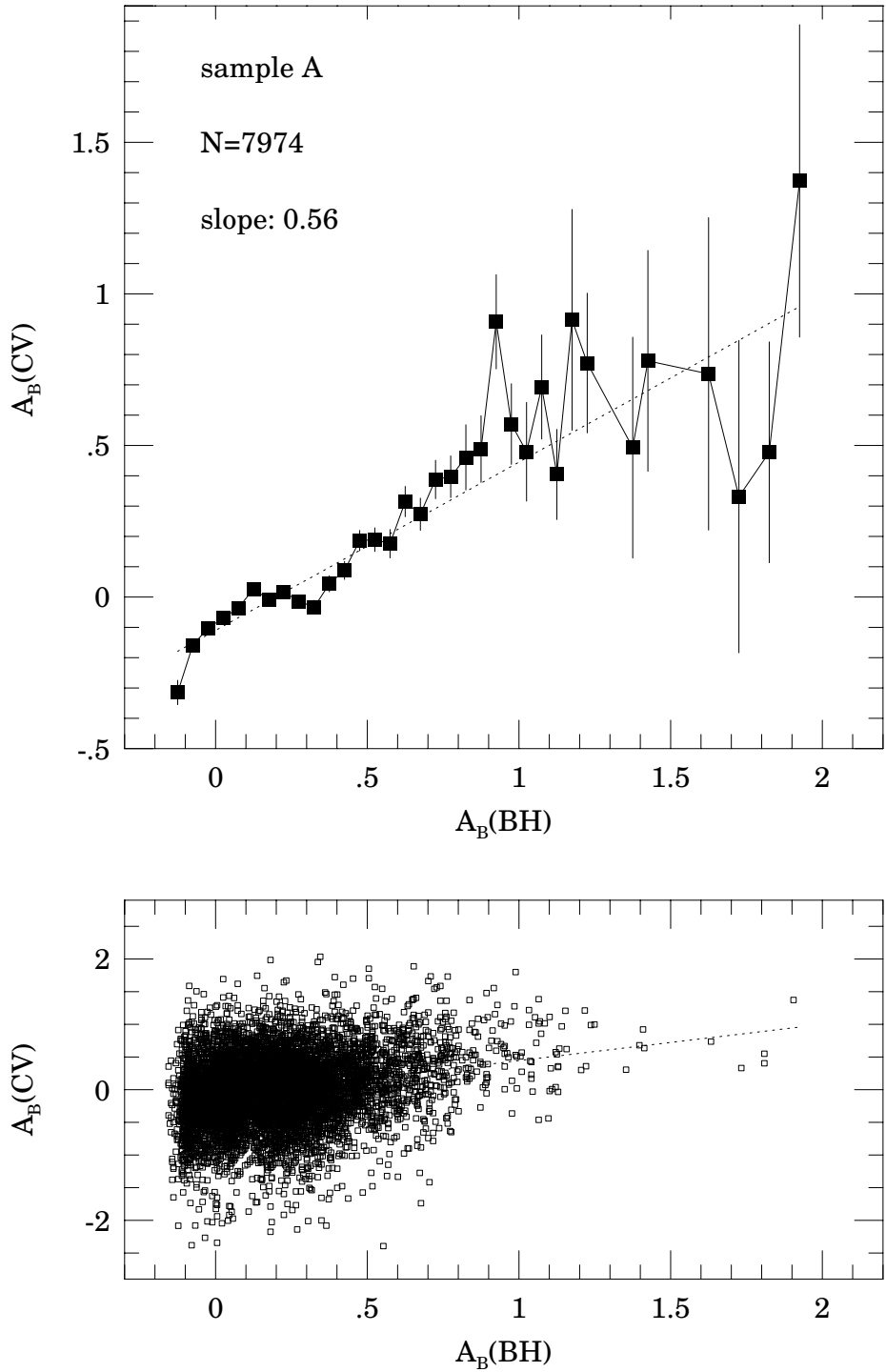}
 \caption{
The same as Fig. 5 but for extinction of BH $A_B(BH)$.
}
\end{figure*}

\clearpage
\begin{figure*}
\vspace{220mm}
 \includegraphics{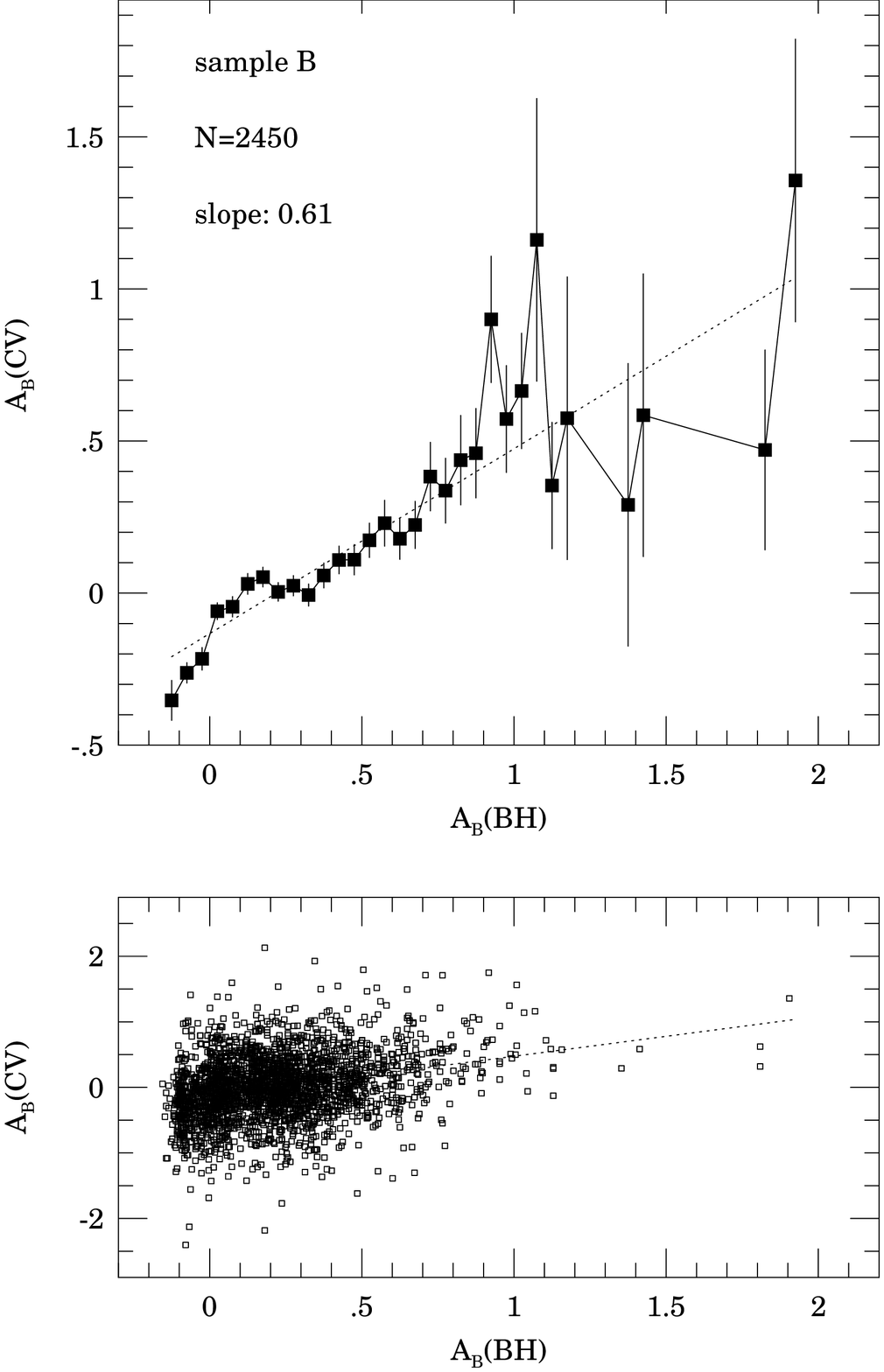}
 \caption{
The same as Fig. 7 but for sample "B".
}
\end{figure*}

\clearpage
\begin{figure*}
\vspace{220mm}
 \includegraphics{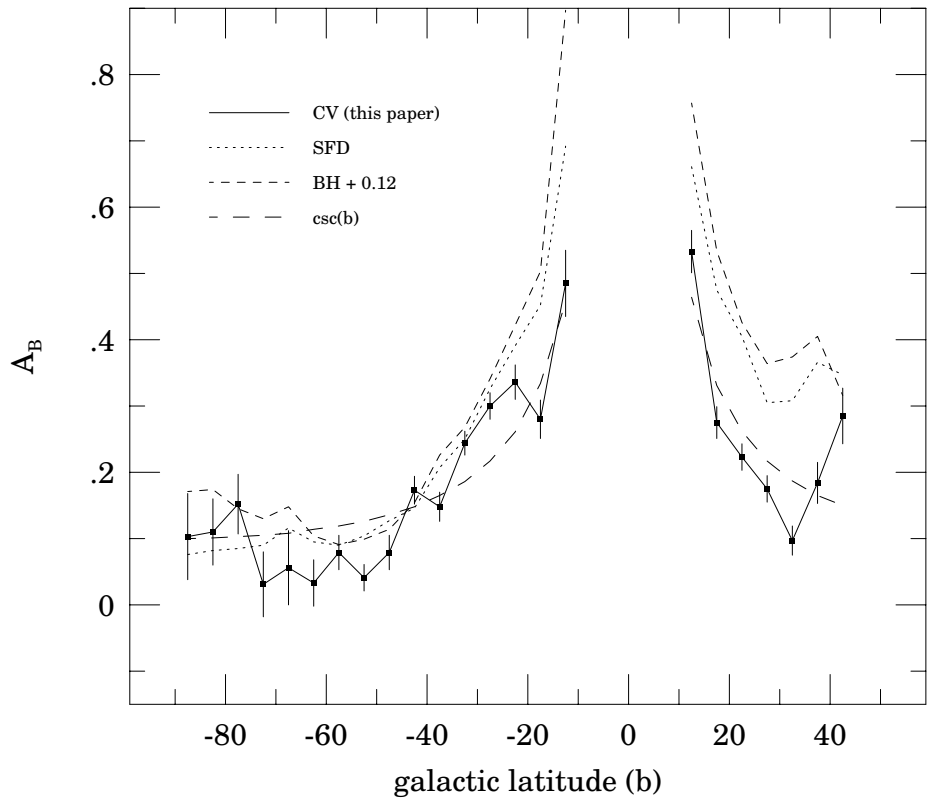}
 \caption{
Galactic latitude dependence of our absolute extinction
(represented by the solid
broken line with $1\sigma$ error bars) compared with the extinction
according to SFD, BH and the cosecans law. The BH extinction have been
corrected by adding 0.12 magnitude constant. 
Note north-south assymetry of our extinction near the 
Galactic equator ($|b|<40^o$) and non-zero extinction
near the Galactic south pole ($b \approx -90$).
}
\end{figure*}

\begin{figure*}
\vspace{220mm}
 \caption{
 \includegraphics{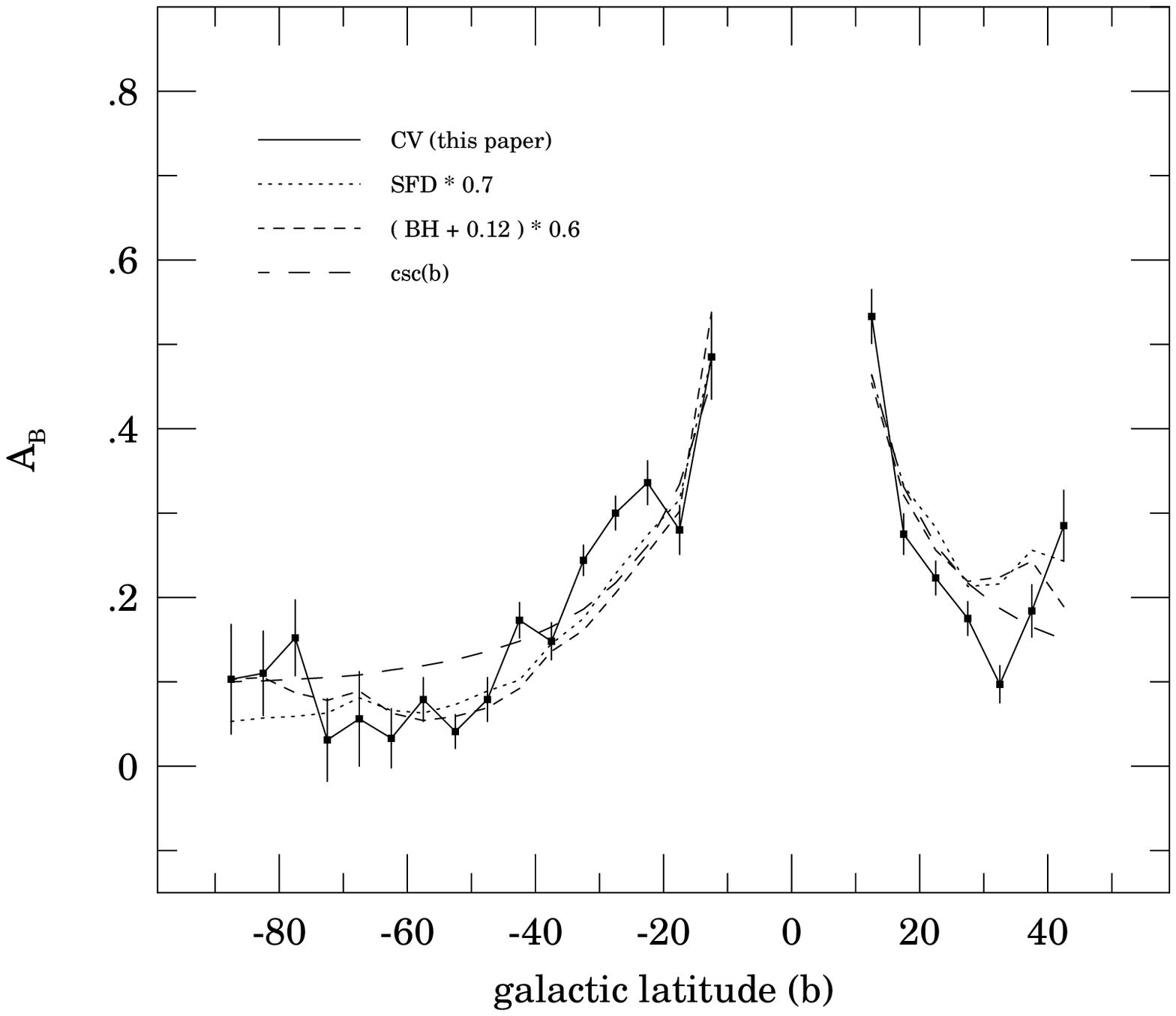}
The same as Fig. 9 but with SFD extinction multiplied by the factor 
0.7 and for corrected BH extinction multiplied by 0.6.
}
\end{figure*}


\begin{thebibliography}{99}
\bibitem{b10} Arce, H. G., Goodman, A. A. 1999a, ApJ, 512, L135
\bibitem{b11} Arce, H. G., Goodman, A. A. 1999b, ApJ, 517, 264
\bibitem{b13} von Braun, K., Mateo, M. 2001, AJ, 121, 1522
\bibitem{b15} Burstein, D., Heiles, C. 1978, ApJ, 225, 40 (BH)
\bibitem{b16} Burstein, D., Heiles, C. 1982, AJ, 87, 1165
\bibitem{b20} Cho{\l}oniewski, J, Valentijn, E. A. 1991, Messenger, 63, 1
\bibitem{b20} Cho{\l}oniewski, J, Valentijn, E. A. 2003, astro-ph/0309750 (CV)
\bibitem{b25} Fisz, M. 1963, Probability Theory and Mathematical
Statistics. John Wiley \& Sons, New York, London
\bibitem{b27} Gonzalez, R. A., Fruchter, A. S., Dirsch, B. 1999, ApJ,
  515, 69
\bibitem{b28} Ivans, I. I., Sneden, C., Kraft, P., Suntzeff, N. B., Smith, V. V.,
  Langer, G. E. , Fulbright, J. P. 1999, AJ, 118, 1273
\bibitem{b30} Lauberts, A., Valentijn, E. A. 1989, {\it The
  Surface Photometry Catalogue of the ESO--Uppsala Galaxies,}
  European Southern Observatory, Garching 
\bibitem{b40} Press, W. H., Teukolsky, S. A., Vetterling, W. T., 
  Flannery, B. P. 1992, Numerical Recipes, Cambridge University
  Press
\bibitem{b55} Schlegel, D. J., Finkbeiner, D. P., Davis M. 1998, ApJ,
  500, 525 (SFD)
\bibitem{b70} Stanek, K. Z. 1998a, astro-ph/9802093
\bibitem{b71} Stanek, K. Z. 1998b, astro-ph/9802307

\end{thebibliography}
\end{document}